\documentstyle[seceq,epsf,twoside]{ptptex}
\def\PRL{\em Phys. Rev. Lett.}
\def\PRD{{\em Phys. Rev.} D}
\def\ZPC{{\em Z. Phys.} C}

\def\be{\begin{equation}}
\def\ee{\end{equation}}
\def\bea{\begin{eqnarray}}
\def\eea{\end{eqnarray}}

\def \be {\begin{equation}}
\def \ba {\begin{eqnarray}}
\def \ee {\end{equation}}
\def \ea {\end{eqnarray}}

\def \uu {$u\bar u$}

\def \ss {$s\bar s$}
\def \qq {$q\bar q$}

\def \sig {$\sigma$}

\def \be {\begin{equation}}
\def \ba {\begin{eqnarray}}
\def \ee {\end{equation}}
\def \ea {\end{eqnarray}}

\def\fz{$f_0(980)$}
\def\az{$a_0(980)$}
\def\Kz{$K_0^*(1430)$}
\def\fzz{$f_0(1300)$}

\def\azz{$a_0(1450)$}
\def\ss{$ s\bar s $}
\def\uu{$u\bar u+d\bar d$}
\def\qq{$q\bar q$}
\def\KK{$K\bar K$}
\def\sig{$\sigma$}
\def\lsim{\;\raise0.3ex\hbox{$<$\kern-0.75em\raise-1.1ex\hbox{$\sim$}}\;}
\def\gsim{\raise0.3ex\hbox{$>$\kern-0.75em\raise-1.1ex\hbox{$\sim$}}}
\notypesetlogo  
\title{How to understand the lightest scalars\footnote{Invited talk at the Conference: "Possible Existence of the Light $\sigma$ Resonance and
its Implications to Hadron Physics", Yukawa Institute for Theoretical Physics, Kyoto, Japan, 11-14th of June 2000.}}
\author{%
Nils A. T\"ornqvist\footnote{email: Nils.Tornqvist@Helsinki.fi} }
\inst{Physics Department, 
POB 9, FIN--00014, University of Helsinki, Finland}
\date{August 1, 2000}
\abst{Based on previous papers I discuss how I understand the lightest scalars
using a general coupled channel model. This model
includes all light two-pseudoscalar  thresholds, constraints from Adler
zeroes, flavour symmetric couplings, 
unitarity and physically acceptable analyticity. 
One finds that with a large 
coupling there can appear  two physical resonance poles on the second sheet although only 
one bare quark-antiquark
state is put in. The f0(980) and f0(1370) resonance poles are thus 
in this model two manifestations of the same strange-antistrange quark  state. 
On the other hand, the isoscalar state containing u and d quarks becomes (when unitarized and strongly distorted 
by hadronic mass shifts) a very broad resonance, with its
pole at 470-i250 MeV. This is the sigma meson
required by models for spontaneous breaking of chiral symmetry.
}

\begin{document}
\maketitle
\setcounter{tocdepth}{4}
\section{ Introduction}
This talk is based on my previous papers~\cite{NAT,NAT2,NAT3,NAT4}, 
including a few new comments. I try to explain how one can understand the controversial light scalar mesons, and show that one can describe the S-wave data on the
light   \qq\ nonet with   a model which includes most well established
theoretical constraints: 
\begin{itemize}
\item Adler zeroes as required by chiral symmetry,
\item all light two-pseudoscalar (PP) thresholds with flavor symmetric couplings in a coupled channel framework
\item physically acceptable analyticity, and 
\item unitarity.
\end{itemize}  A unique
feature of this model is that it simultaneously describes the  whole scalar
nonet and one obtains a good representation of a large
set of relevant data. Only six parameters,
which all have a  clear physical interpretation, are needed: 
an overall coupling constant
($\gamma=1.14$), the bare mass of the $u\bar u$ or $d\bar d$ state,
the extra mass for a strange quark ($m_s-m_u=100$ MeV),
a cutoff parameter ($k_0=0.56$ GeV/c), an Adler zero parameter for $K\pi$, and a phenomenological 
parameter enhancing the $\eta\eta '$ couplings. 
\section{ Understanding the S-waves}
In Figs.~1-3 we show the obtained fits to the $K\pi$, $\pi\pi$ S-waves and
to the  \az\ resonance peak in $\pi\eta$. The partial 
wave amplitude is in the case of one 
\qq\ resonance, such as the \az\ can be written 
\be
A(s)=-Im\Pi_{\pi\eta}(s)/[m_0^2+Re\Pi (s)-s +iIm\Pi (s)], \label{PWA}
\ee
\noindent where
\bea
Im\Pi (s)&=&\sum_i Im\Pi_i(s)  \label{impi}\\
       &=&-\sum_i \gamma_i^2(s-s_{A,i})\frac{k_i}{\sqrt s}e^{-k_i^2/k_0^2}
           \theta(s-s_{th,i})\ ,\nonumber \\
Re\Pi_i(s)&=&\frac 1\pi{\cal P}\int^\infty_{s_{th,1}} \frac{Im\Pi_i (s)}{s'-s} ds'
\ . \label{repi}
\eea
Here the coupling constants $\gamma_i$ are related by flavour
symmetry and the OZI rule, such that there is only one over all parameter
$\gamma$. The $s_{A,i}$ are the positions of the Adler
zeroes, which normally are  $s_{A,i}=0$, except  $s_{A,\pi\pi}=m_\pi^2/2$, and  
$s_{A,K\pi}$, which is a free parameter.

 In the flavourless channels 
the situation is a little more complicated than eqs. (1-3)
since one has both \uu\ and \ss\  states, requiring a two dimensional 
mass matrix (See Ref.~\cite{NAT}).
 Note that the sum runs over all light
PP thresholds, which means three for the \az : $\pi\eta ,\ K\bar K,\pi\eta'$
and three for the 
\Kz : $ K\pi ,\ K\eta ,\ K\eta' $, while for the $f_0$'s there are
five channels: $\pi\pi ,K\bar K ,\ \eta\eta ,\ \eta\eta' ,\ \eta'\eta'$.
Five channels means the amplitudes have $2^5=32$! different Riemann sheets, 
and in principle there can be poles on each of these sheets.
In Fig.~4 we show as an example the running mass, $m_0^2+Re\Pi(s)$,
 and the width-like function,
$Im\Pi(s)$, for the I=1 channel. The crossing point of the running mass with
$s$ gives the $90^\circ$  mass of the \az .
 The magnitude of the \KK\ component in the \az\ is determined by
$-\frac d{ds}Re\Pi(s)$ which is large in the resonance region just
below the \KK\ threshold. These functions fix the
PWA of eq.(1) and Fig.~3. In Fig. 5 the running mass and width-like function
for the strange channel are shown. These fix the shape of the $K\pi$ phase
shift and absorption parameters in Fig. 1. 

Four out of our six parameters are fixed by the $K\pi$ data leaving only
$m_s-m_u=100$ MeV to "predict" the \az\ structure Fig.~3, and the
parameter $\beta$  to get the $\pi\pi$ phase shift right 
above 1 GeV/c. 
 One could discard the $\beta$
parameter if one also included the next group of important
thresholds or pseudoscalar ($0^{-+}$) -axial ($1^{+-}$)
 thresholds, since then the $K\bar K_{1B}+c.c.$
thresholds give a very similar contribution to the mass matrix as $\eta\eta'$.
As can be seen from Figs.~1-3 the model gives a good
description of the relevant  data.                                              
\begin{table}[t] 
\caption{Resonances in the S-wave  $ PP \to PP $ amplitudes~$^1$ 
The first resonance is the \sig\
which we name here $f_0(\approx 500)$. The two following are both
manifestations of the same $s\bar s$ state. The \fz\ and \az\ have no
approximate Breit Wigner-like description, and the $\Gamma_{BW}$ given for
\az\ is rather the peak width. 
The mixing angle $\delta_S $ for the
$f_0(\approx 500)$ or $\sigma$  is with respect to
$u\bar u +d\bar d$, while for the two heavier
$f_0$'s it is with respect to $s\bar s$. }
\vspace{ 0.5cm} 
\begin{center}
\begin{tabular}{|c|c|c|c|l|}
\hline
resonance&$m_{BW}$&$\Gamma_{BW}$&$\delta_{S,BW}$&Comment\\
\hline
$f_0(\approx 500)$&860&880&$(-9+i8.5)^\circ$&The $\sigma$ meson.\\
$f_0(980)$&-&-&-&First near \ss\ state\\
$f_0(1300)$&1186&360&$(-32+i1)^\circ$&Second
near \ss\ state\\
$K_0^*(1430)$&1349&498&-&The $s\bar d$ state\\
$a_0(980)$&987&$\approx$100&-&First I=1 state\\
\hline
\end{tabular}
\vspace{0.5cm}\end{center}
\end{table}
\begin{table}[t] \caption{
The pole positions of the same resonances as in Table 1.
The last entry is an image pole of the
\az , which in an improved fit could represent the $a_0(1450)$.
The \fzz\ and \Kz\ poles appear simultaneously
on two sheets since the $\eta\eta$ and the $K\eta$ couplings, respectively,
 nearly vanish. The mixing angle $\delta_S $ for the
$f_0(\approx 500)$ or $\sigma$  is with respect to
$u\bar u +d\bar d$, while for the two heavier
$f_0$'s it is with respect to $s\bar s$. }
\vspace{ 0.5cm} 
\begin{center}
\begin{tabular}{|c|c|c|c|c|c|}
\hline
resonance&$s_{\rm pole}^{1/2}$& $[{\rm Re}s_{pole}]^{1/2}$&
$\frac{-{\rm Im}\ s_{pole}}{m_{pole}}$&$\delta_{S,pole}$&Sheet\\
\hline
$f_0(\approx 500)   $&$ 470-i250$& 397& 590 &$(-3.4+i1.5)^\circ $&II    \\
$f_0(980)   $&$1006-i17 $&1006& 34  &$(0.4+i39)^\circ   $&II    \\
$f_0(1300)  $&$1214-i168$&1202& 338 &$(-36+i2)^\circ    $&III,V \\
$K_0^*(1430)$&$1450-i160$&1441& 320 & -                  &II,III\\
$a_0(980)   $&$1094-i145$&1084& 270 & -                  &II    \\
$a_0(1450)? $&$1592-i284$&1566& 578 & -                  &III   \\
\hline
\end{tabular}
\vspace{0.5cm}\end{center}
\end{table}
In Ref.~\cite{NAT} I did not look 
for the broad and light sigma on the second sheet in my model, since I looked for only those 
 poles which are nearest to the
physical region, and which  could complete
the light \qq\ scalar nonet.    I found parameters for these close to the
conventional lightest scalars in the PDG tables for the \fz , \fzz , \az\ and \Kz . Only a little later I realized with Roos\cite{NAT2} that I had missed the \sig , and that both my \fz\ and \fzz , in fact, originated from the same \ss\ input bare state.
 \section{One $q\bar q$ pole can give rise to two resonances}
As pointed out by Morgan and Pennington~\cite{morgan} for each \qq\ state 
there are, in general, apart from the nearest pole also  image poles, usually
located far from the physical region.
As explained in more detail in Ref.~\cite{NAT2}
 some of these can (for a large enough coupling and sufficiently heavy threshold) come so close to the physical region that they make new resonances.     
And, in fact, there are more than four physical poles  with different isospin,
in the output spectrum of my model, although only four bare states are put in!
In Table 2 I list the significant  pole positions.

All these poles are manifestations of {\it the same nonet}~\cite{NAT2}.
The \fz\ and the \fzz\ turn out to be  two 
manifestations of the same \ss\ state.  
(See Ref\cite{NAT2} for details).
There can be two crossings with the running mass $m_0^2+{\rm Re}\Pi(s)$, one near the threshold
and another at higher mass, and each one is related to a different pole at the second sheet
(or if the coupling is strong enough the lower one could even become a bound state pole, below the threshold, on the first sheet).

 Similarily the \az\ and the  \azz\ could be two manifestations of the $u\bar d$ state. 
After I had realized this I, of course, had to find the \uu\ pole of my model. Then I found my 
light and broad \sig . 

\section{The light $\sigma$ resonance}

A light scalar-isoscalar meson (the \sig ), with a mass of
 twice the constituent  $u,d$ quark mass coupling strongly to
$\pi\pi$ is of importance in many models for spontaneous breaking
of chiral symmetry, and for the understanding of all hadron masses, since the \sig\
generates  constituent quark masses  different from  chiral quark masses.
 Thus most of the nucleon mass can be generated by its coupling to
the $\sigma$, which acts like an effective Higgs-like 
 boson for the hadron spectrum.  However, when I worked on this model the lightest well established mesons in the Review of Particle Physics did not include the \sig . 
The lightest isoscalars at that time, which had
the quantum numbers of the $\sigma$ were the \fz\ and \fzz . These do
not have the right properties being  both too narrow. Furthermore, \fz\ couples mainly
to $K\bar K$, and \fzz\ is too heavy.

Thus I found it exciting that the  important pole in my \uu\ channel turned out to be  the the first pole in Table 2,
which I beleve is  the long sought for \sig =$f_0(\approx 600)$ with twice the constituent
quark mass as in the famous Nambu relation ($m_\sigma\approx 2m_q$).
It gives rise to a very broad 
Breit-Wigner-like background, dominating 
 $\pi\pi$ amplitudes below 900 MeV.
It has the right mass and width and large $\pi\pi$
coupling as predicted by the \sig\ model.

The existence of this meson becomes evident if one studies the \uu\ channel
separately. This can be done within the model, perserving unitarity and
analyticity, by letting the $s$ quark (and $K$, $\eta$ etc.) mass
go to infinity. Thereby one  
eliminates the influence from \ss\ and \KK\ channels, which perturb $\pi\pi$
scattering very little through mixing below 900 MeV. 
The \uu\ channel is then seen to be dominated by the sigma below 900 MeV. 
For more details see Ref.\cite{NAT4}.

Isgur and Speth~\cite{Isgur}  criticised this result 
claiming that details of crossed channel
exchanges, in particular $\rho$ exchange, are important. In our reply~\cite{Isgur}
to this we emphasized the well known result from dual models, 
that a sum of $s$-channel resonances also describes $t$-channel phenomena.
 In my model the crossed channel singularities are represented, although 
in a very crude way, through 
the form factor $F(s)$, which in an N/D language is related to the N function. 
See also Igi and Hikasa\cite{igi}. Improvements to the model can be done by allowing
for a more complicated analytic form for $F(s)$. 

I believe more important than the details of crossed channel singularities is
the fact that I included the chiral (Adler) zeroes. These were absent in my first
attempt in 1982 to fit the scalars using a unitarized quark model\cite{NAT}.

\section{The large mass difference between the $K^*_0(1430)$ and the $a_0(980)$ }
Many authors argue that the $a_0(980)$ and $f_0(980)$ are not \qq\ states,
since in addition to being very close to the \KK\ threshold, they are much lighter
than the first strange scalar, the  $K^*_0(1430)$.
Naively one expects a mass difference between the strange and nonstrange meson to be of the order of the strange-nonstrange quark mass difference, or a little over 100 MeV. This is also one 
of the reasons why some authors want to have a 
lighter strange meson, the $\kappa$, near 800 MeV. Cherry and Pennington\cite{cherry} recently have
strongly argued against its existence. 

Figs. 4 and 5 explain why  one can easily understand this large mass splitting as a secondary effect of the
large pseudoscalar mass splittings, and because of the large mass shifts coming from
the loop diagrams involving the PP thresholds. If one puts Figs. 4 and 5 on top of
each other one  sees that  the 3 thresholds $\pi\eta,\ K\bar K,\ \pi\eta$ all lie
relatively close to the $a_0(980)$, and all 3 contribute to a large mass shift.
On the other hand, for the $K^*_0(1430)$ the $SU3_f$ related thesholds ($K\pi,\ K\eta'$) lie far 
apart from the $K^*_0$, while the $K\eta$ nearly decouples because of the physical value of the pseudoscalar mixing angle.

\section{Concluding remarks}
An often raised question is: Why are the mass shifts required by unitarity
so much more important for the scalars than, say, for the vector  mesons?
The answer is very simple, and there are two main reasons: 
\begin{itemize}
\item 
The scalar coupling to two pseudoscalars is very much larger
than the corresponding coupling for the vectors, both experimentally and theoretically
(e.g. spin counting gives  3 for the ratio of the two squared couplings).
\item For the scalars the thresholds are S-waves, giving nonlinear square root cusps in
the $\Pi(s)$ function, wheras for the vectors the thresholds are P-waves, giving a smooth
$k^3$ angular momentum and phase space factor. 
\end{itemize}
\vskip .2cm

One could argue that the two states \fz\ and \az\ are a kind of
$K\bar K$ bound states (c.f. Ref.~\cite{wein}), 
since these have a large component of $K\bar K$
in their wave functions. However, the dynamics of these states is quite 
different from that of normal two-hadron bound states.
If one wants to consider them as \KK\ bound states,
it is the $K\bar K \to s\bar s \to K\bar K$ interaction which  creates their
binding energy, not the hyperfine interaction as in Ref.~\cite{wein}. Thus, although they may spend most of their 
time as $K\bar K$, they 
owe their existence to the \ss\ state. Therefore, it is more natural to consider
the \fz\ and \fzz\ as two manifestations of the same \ss\ state. 

The wave function of the $a_0(980)$ (and $f_0(980)$) can be pictured as a relatively small core
of \qq\ of typical \qq\ meson size (0.6fm), which is  surrounded by a much larger standing 
S-wave of \KK . This picture also gives a physical explanation of the narrow width: In order to decay
to $\pi\eta$ the \KK\ component must first virtually annihilate near the origin to \qq . Then
the \qq\ can decay to $\pi\eta$ as an OZI allowed decay.




\twocolumn[\hsize\textwidth\columnwidth\hsize\csname @twocolumnfalse\endcsname
]
\begin{figure}
\epsfxsize=7 cm
\epsfysize=6. cm
\epsffile{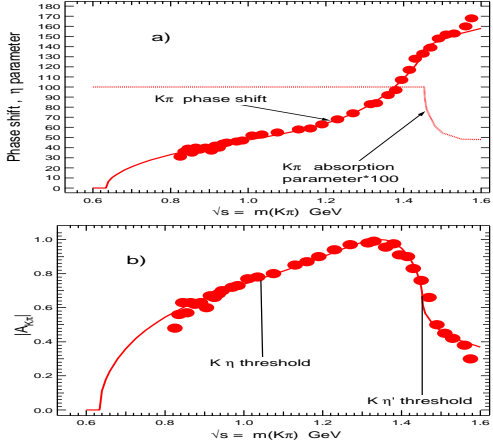}
\caption{ (a) The
$K\pi$ S-wave
phase shift and (b) the magnitude of the $K\pi$
partial wave amplitude compared with the model predictions, which
fix 4 ($\gamma$, $m_0+m_s$, $k_0$ and $s_{A,K\pi}$) of the 6
parameters.}
\end{figure}
\begin{figure}
\epsfxsize=5.3 cm
\epsfysize=7. cm
\hskip 0.8cm
\epsffile{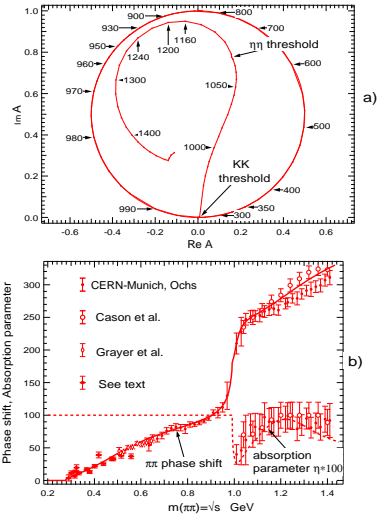}
\caption{ (a) The $\pi\pi$ Argand diagram and (b) phase shift predictions
are compared with data. Note that most of the parameters were fixed by the
data in Fig.1. For more details see Ref.~$^{1,2}$. }
\end{figure}

\begin{figure}
\epsfxsize=5. cm
\epsfysize=6 cm
\hskip .8 cm \epsffile{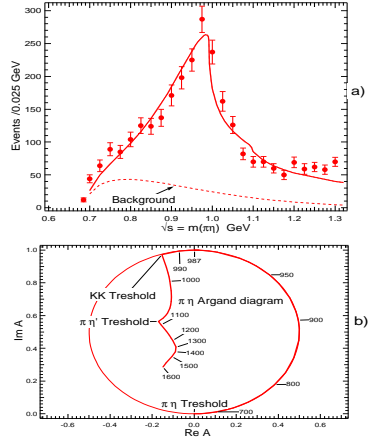}
\caption{ (a) The $a_0(980)$ peak compared with model prediction and (b) the 
predicted $\pi\eta$ Argand diagram}
\end{figure}

\begin{figure}
\epsfxsize=4.7 cm
\epsfysize=3 cm
\hskip 0.9cm
\epsffile{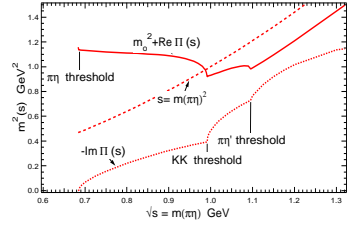}
\caption{ The running mass $m_0+{\rm Re}\Pi(s)$ and Im$\Pi (s)$ of the $a_0(980)$.
The strongly dropping running mass at 
the $a_0(980)$ position, below the $K\bar K$ 
threshold contributes to the narrow shape of the peak in Fig. 3a.}
\end{figure}
                                  
\begin{figure}
\epsfxsize=7 cm
\epsfysize=4 cm
\epsffile{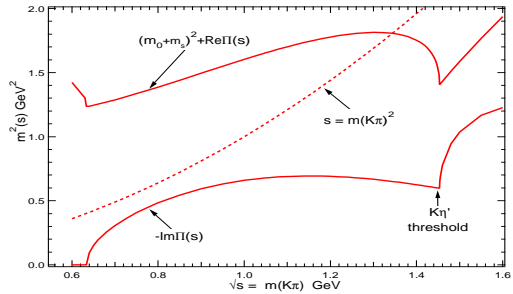}
\caption{The running mass and width-like function Im$\Pi(s)$ for the
$K^*_0(1430)$. The crossing of $s$ with the running mass gives the 90$^\circ$
phase shift mass, which roughly corresponds to a naive Breit-Wigner mass,
where the running mass is put constant.  }
\end{figure}


\end{document}